\documentclass[12pt]{iopart}
\usepackage[dvips]{graphicx}
\usepackage{supercite}

\begin{document}

\sloppy

\title[Electrophoretic mobility of a charged colloidal particle]
{Electrophoretic mobility of a charged colloidal particle:
A computer simulation study}
\author{Vladimir Lobaskin\footnote[1]{To whom correspondence should be addressed
(lobaskin@mpip-mainz.mpg.de)}, Burkhard D\"unweg, and Christian Holm}

\address{Max Planck Institute for Polymer Research,
Ackermannweg 10,
D-55128 Mainz, Germany}
\date{\today}
\begin{abstract}
  We study the mobility of a charged colloidal particle in a constant
  homogeneous electric field by means of computer simulations. The
  simulation method combines a lattice Boltzmann scheme for the fluid
  with standard Langevin dynamics for the colloidal particle, which is
  built up from a net of bonded particles forming the surface of the
  colloid. The coupling between the two subsystems is introduced via
  friction forces. In addition explicit counterions, also coupled to
  the fluid, are present. We observe a non-monotonous dependence of the
  electrophoretic mobility on the bare colloidal charge. At low
  surface charge density we observe a linear increase of the mobility
  with bare charge, whereas at higher charges, where more than half of
  the ions are co-moving with the colloid, the mobility decreases with
  increasing bare charge.
\end{abstract}

\submitto{\JPCM}
\maketitle

\section{Introduction}
\label{sec:intro}

Many important properties of colloidal dispersions are directly or
indirectly determined by the electric charge of the colloidal particles. The
charge and the associated electrostatic potential normally yield a
repulsive interaction between the particles that stabilizes the
dispersion. The current understanding of electro--rheological
phenomena is based on a central quantity, the electrokinetic potential
or zeta ($\zeta$) potential, which is decisive for e.~g. the
electrophoretic motion of the particles, the overall flow behavior of
the dispersion, sedimentation, and flocculation. Calculating both the
electrostatic and the electrokinetic potential, and establishing a
relation between them, is however a challenging task, which has
attracted the attention of colloid scientists for decades
\cite{Smoluchowski,Hueckel,Dukhin,Hunter}. For the
electrokinetic potential, the main obstacle is the strong coupling
between the electric and hydrodynamic degrees of freedom. Partial
decoupling is only possible in the case of weak external fields or for
unperturbed double layers.

In a constant and homogeneous electric field a charged particle
accelerates until the friction force, which is proportional to the
velocity, equals the electrostatic force. The electrophoretic mobility
$\mu$ of a charged particle is defined as the ratio between its
stationary velocity $v_{c}$ and the applied electric field $E$,
\begin{equation}
\mu = v_{c} / E   .   
\label{eq:1}
\end{equation}

In the classical theories, it is assumed that one can define a
so--called shear plane (or slip surface), which divides the
surrounding fluid into two parts: The inner part that moves with the
particle, due to strong electrostatic and hydrodynamic coupling of the
electric double layer to the colloidal surface, and the outer part
which typically moves in the opposite direction, due to the current of
the counterions. The zeta potential is defined as the electrostatic
potential at the slip surface, and therefore in general differs from
the potential at the particle surface. Once this potential is known,
the remaining calculation becomes more or less straightforward, if
still spherical symmetry is assumed. The electrophoretic mobility
$\mu$ can be expressed in general through the $\zeta$ potential as
\begin{equation}
\mu = \frac{\varepsilon \zeta}{ 6 \pi \eta} f(\kappa R)
\label{eq:mu}
\end{equation}
where $\eta$ is the solvent viscosity, $\varepsilon = 4 \pi
\varepsilon_0 \varepsilon_r$ is the dielectric constant of the fluid,
composed of the vacuum and relative dielectric constant,
$\varepsilon_0$ and $\varepsilon_r$, respectively, $\kappa^{-1}$ is the
Debye screening length, $R$ denotes the colloidal radius, and $f$ is a
function of the salt concentration and colloidal radius. For small
colloids and~/~or low electrolyte concentrations $\kappa R \rightarrow
0$ and $f(\kappa R) \rightarrow 1$, and the H\"uckel-Onsager limit is
recovered \cite{attard00a}, whereas for $\kappa R \rightarrow \infty$
one obtains $f(\kappa R) \rightarrow 3/2$, which is known as the
Helmholtz-Smoluchowski limit. At intermediate values of $\kappa R$,
the values of $f(\kappa R)$ depend on the applied theory
\cite{henry31a,wiersema66a,obrien78a,lozada01a}.

The electrophoretic mobility is easily accessible in experiment and represents
a valuable source of information for a colloid scientist \cite{Dukhin,Hunter}.
The desired microscopic information about the actual size of the particles and
their surface potential is however not straightforward to extract. The
main reason is that the experimental data only provide information about the
motion of the particles together with their double layer. The structure of the
latter, however, is not easy to calculate analytically, due to the strong
electrostatic coupling and the influence of hydrodynamics, such that linear
theories will not work, while different analytical approximation schemes give
different answers.

A computer simulation including both electric charges and hydrodynamic
interactions provides a unique chance to examine the double layer
structure and various dynamic properties such as the particle mobility
or the conductivity at the same time. There has been a recent
Molecular Dynamics (MD) simulation study where the solvent particles
were modeled explicitly \cite{tanaka01a}. However the authors were
more interested in the overcharging effects occurring at high Coulomb
coupling. On the other hand, a significant effort had been made to
include hydrodynamic effects in a colloidal simulation on a less
expensive level, namely via the lattice Boltzmann (LB) method
\cite{Ladd0,Ladd4,Frenkel1,Horbach1}. This latter approach models the
colloidal particles as extended hollow spheres, while stick boundary
conditions at the surface are implemented via bounce-back collision
rules. The method was extended to charged colloids by Horbach and
Frenkel \cite{Horbach1}. In their work, the small ions were
represented by LB populations, such that the electrostatics is
essentially taken into account on the level of the Poisson-Boltzmann
equation. Unfortunately, this method has two disadvantages we would
like to avoid: Firstly, the discrete nature of the ions and
correlations beyond the Poisson-Boltzmann level are not taken into
account; secondly, one cannot avoid a leakage of charge into (and out
of) the sphere, such that it is hard (if not impossible) to maintain a
well-defined Debye layer of charges around it \cite{Horbach2}.

For these reasons, we rather prefer to simulate charge-stabilized
colloidal dispersions with explicit counterions; this automatically
eliminates both problems. The disadvantage (with which we have to
live) is however that only a limited range of size ratios between
colloidal particle and counterions is accessible. Recently we proposed
a suitable hybrid model which implements this philosophy, while
keeping the successful representation of hydrodynamics in terms of the
LB approach \cite{NJP}. The solvent is modeled via LB, while MD is
done for the solute (colloidal particles and counterions). However, in
contrast to the bounce-back implementation of the coupling, we rather
use point particles (``monomers''). It should be noted that this term
refers only to the coupling to the solvent, while the monomers do have
a finite size with respect to their mutual interaction. The coupling
to the solvent is inherently dissipative in nature: Each monomer is
assigned a phenomenological friction coefficient, and the friction
force between solvent and particle is proportional to the relative
velocity. The flow velocity at the location of the monomer is obtained
via linear interpolation from the surrounding lattice sites, and this
implies that the lattice spacing should be of the order of the monomer
size. The advantage of this approach is that it is quite flexible,
because various large objects of soft matter physics (colloidal
particles, polymer chains, membranes, etc.) can be built up from
elementary monomers, without major restructuring of the underlying
simulation program. However, it is not
possible to model the colloidal particle just in terms of a single
monomer. While the electrostatic interactions plus the excluded volume
between colloidal particle and counterions could be easily represented
in terms of a single strongly charged particle with large repulsion
radius, such a particle would have inappropriate hydrodynamic
properties. The fluid would be coupled only to the center of the
particle, while a faithful representation of the hydrodynamics
requires stick boundaries at the surface of the particle, or a good
approximation thereof. Furthermore, we wish to faithfully represent
the particle's rotational motion. For these reasons, we add a
two-dimensional tethered network of monomers which we wrap around the
surface of the central sphere, such that the overall structure
resembles a raspberry. The coupling to the solvent is then done only
via the surface sites in terms of the monomer friction
coefficient. For reasonably large friction, this is an excellent
approximation to a stick boundary condition, as has been shown in
Ref. \cite{NJP} in terms of both the translational and the rotational
motion of the sphere.

In the present work, we analyze basic electrokinetic properties of
such a model colloid, where we restrict ourselves to the case of a
single sphere, and a rather elementary analysis. The remainder of this
article is organized as follows: In Sec. \ref{sec:model}, we describe
our simulation model, while Sec. \ref{sec:results} contains the
numerical results on the double layer structure and particle mobility
in a constant electric field. Finally, Sec. \ref{sec:summary}
concludes with a brief summary.

\section{Model}
\label{sec:model}

Our hybrid simulation method involves two subsystems: the solvent that
is modeled via LB with fluctuating stress tensor (i.~e. we run a
constant-temperature version of the LB method) and a Langevin MD
simulation for the particles immersed in the solvent. The LB
simulation is performed using the 18-velocity model \cite{Ladd1},
while the solute is coupled dissipatively to the solvent as described
in \cite{Ahlrichs1,Ahlrichs2}. The fluid simulation consists of
collision and propagation steps, the former being performed with
inclusion of the momentum transfer from the solute particles (surface
beads and ions).

\begin{figure}
\begin{center}
\vskip 0.1in
\includegraphics[clip,width=0.5\textwidth]{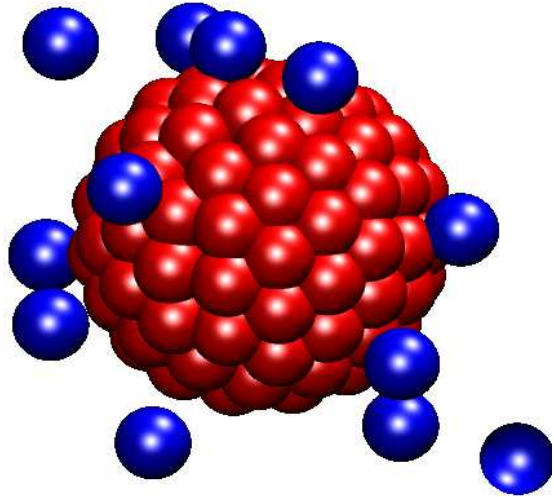}
\end{center}
\caption{
Raspberry-like model of a colloidal sphere. There is a central large 
bead of radius $R=3$ and charge $Z=20$. The small beads of radius 1
are connected with their nearest neighbors on the surface via FENE
bonds. A repulsive soft-core potential is also operating between all
the monomers. The counterions are moving freely in space and interact
with the central bead via the Coulomb potential and the repulsive LJ 
potential. 
}
\label{fig:sphere}
\end{figure}

The colloidal particle is represented by a two-dimensional tethered
bead-spring network consisting of $100$ beads, which is wrapped around
a ball of a radius $\sigma_{cs}$ (for notation, see below), so that
the whole construction resembles a raspberry (see
Fig. \ref{fig:sphere}). The network connectivity is maintained via
finitely extendible nonlinear elastic (FENE) springs,
\begin{equation}
V_{FENE} (r) = - \frac{k R_0^2}{2} \ln
\left(1 - \left( \frac{r}{R_0}\right)^{2} \right) ,
\label{eq:fene}
\end{equation}
where $k$ is the spring constant chosen to be $k=25$, and $R_0$ 
denotes the maximum bond extension, here 1.25.
Furthermore, the beads repel each other by a modified LJ potential
\begin{equation}
V_{LJ} (r) = \left\lbrace \begin{array} {ll} 
4\epsilon_{ij} \left(  \left( \frac{\sigma_{ij}}{r} \right)^{12} 
- \left( \frac{\sigma_{ij}}{r} \right)^{6} + 
\frac{1}{4} \right) & r<2^{1/6} \sigma_{ij}  \\
0  & r\ge 2^{1/6} \sigma_{ij} . \end{array} \right.
\label{eq:lj}
\end{equation}
An additional repulsive LJ bead is introduced at the center of the sphere in
order to maintain its shape. In Eq. \ref{eq:lj}, $i,j$ denote either a central
(``c''), or a surface (``s'') bead, or an ion (``i''). The unit system is
completely defined by the surface bead parameters by setting $\epsilon_{ss}$,
$\sigma_{ss}$, and the surface bead mass $m_s$ to unity. All other beads have
mass 1, too. The interaction between the central bead and the surface beads is
described by $\sigma_{cs} = 3$, which is thus the sphere radius, and
$\epsilon_{cs}=48$, while the interaction with the ions is characterized via
$\sigma_{ci} = 4$ and $\epsilon_{ci}=1$. There are no interactions between the
ions and the surface beads. We place the colloid charge at the central bead
and add an appropriate number of counterions (LJ beads with $\sigma_{ii} = 1$,
$\epsilon_{ii}=1$ ) outside the sphere. The electrostatic interaction is taken
into account via the Coulomb potential
\begin{equation}
V_{el} (r) = \lambda_B k_B T \frac{z_i z_j}{r}
\label{eq:electrostatics}
\end{equation}
between the various charges, where the standard Ewald summation
technique \cite{Allen} is applied, as it is appropriate for a small
number of charges. In Eq. \ref{eq:electrostatics}, $\lambda_B=e^2/
\left(4\pi\varepsilon_0 \varepsilon_r k_B T\right)$ is the Bjerrum length, 
$k_B$ the Boltzmann constant, $z_i$ the valency of species $i$ in
units of the elementary charge $e$, and $T$ the temperature. The
Bjerrum length in all our runs was set to $\lambda_B=1.3$. This value
is motivated by an attempt to mimic an aqueous dispersion of spherical
sodium dodecyl sulfate micelles, which in reality have radius of 2 nm
and carry 60 elementary charges \cite{lobaskin99,lobaskin01}.

The LB lattice constant is chosen as one (in our LJ unit system), and
the fluid is simulated in a cubic box with periodic boundary
conditions and box size $L=30$, which defines the effective finite
concentration of our system; this corresponds to a volume fraction of
about 1\% if $\sigma_{ci}$ is taken as the particle radius. The force
acting on the surface beads (or ions) is given by
\begin{equation}
\vec{F}= - \Gamma \left( \vec{V} - \vec{u} \right) + \vec{f} .
\label{eq:4}
\end{equation}
Here, $\Gamma$ is the ``bare'' \cite{Ahlrichs2} friction coefficient,
$\vec{V}$ and $\vec{u}$ are the velocities of the bead and the fluid
(at the position of the bead), respectively, while $\vec{f}$ is a
Gaussian white noise force with zero mean, whose strength is given via
the standard fluctuation-dissipation theorem
\cite{Ahlrichs1,Ahlrichs2} to keep the surface beads and ions at the
same temperature as the solvent. In our simulation we used a friction
constant $\Gamma = 20$, a temperature $k_B T = 1$, a fluid mass
density $\rho = 0.85$, and a kinematic viscosity $\nu=3$, resulting in
a dynamic viscosity $\eta=2.55$. At least 20000 MD steps were
performed to equilibrate the initial random bead configuration before
the interaction with the LB solvent was turned on. A multiple time
step technique was used, with the MD time step of 0.005 LJ time units
and LB field update interval of 0.01.  Further details on the method
can be found in \cite{NJP,Ahlrichs1,Ahlrichs2}.

We should note that the application of periodic boundary conditions
makes the electrophoresis in our system somewhat different from the
experimental case where a multi-colloid system reacts to the external
field. The particle is both hydrodynamically and electrostatically
coupled to its periodic images. Therefore, we measure in fact the
mobility of a periodic particle array arranged in a simple cubic
lattice with lattice spacing of the simulation box size.

\section{Results}
\label{sec:results}

\begin{figure}
\vskip 0.1in
\begin{center}
\includegraphics[clip, width=0.75\textwidth]{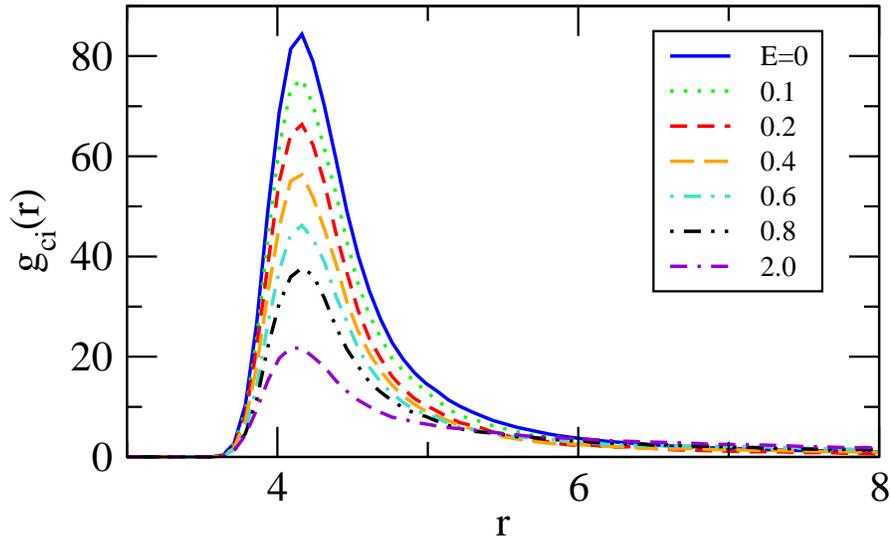}
\end{center}
\caption{
Colloid-ion radial distribution functions for a colloidal particle 
of charge -60 in an external electric field $E$. }
\label{fig:rdf1}
\end{figure}

We first investigated the character of the particle motion and the
ionic density distribution in a constant homogeneous electric
field. We studied a particle of charge -60 with 60 counterions. No
salt ions were added, i.~e. the screening layer consisted of
counterions only. Several field values were considered, beginning in
the regime in which the drift was barely detectable, and ending in the
regime where the mean drift velocity was of order unity in our LJ
units. In all cases, a steady motion with fairly constant velocity
(except for thermal fluctuations) was reached within a few LJ time
units. In Fig. \ref{fig:rdf1} one can see the spherically averaged
stationary state radial distribution functions (rdf's) for the
colloid-ion correlation for several field values. The rdf's are
qualitatively similar for the static and the dynamic case; in essence,
only the peak height of the distribution is affected by the external
field. The main peak of the colloid-ion rdf decreases substantially
upon increasing the external field. This effect can be explained by
(i) a stronger electric force acting on the ions and (ii) faster
motion of the colloid with its ionic cloud and hence stronger
friction, which in combination result in stripping an additional
fraction of the double layer. It is important to note that this latter
regime is already beyond the limits of validity of the standard
electrokinetic theory, which is based on linear response treatment of
the external field effects.

\begin{figure}
\vskip 0.1in
\begin{center}
\includegraphics[clip, width=0.75\textwidth]{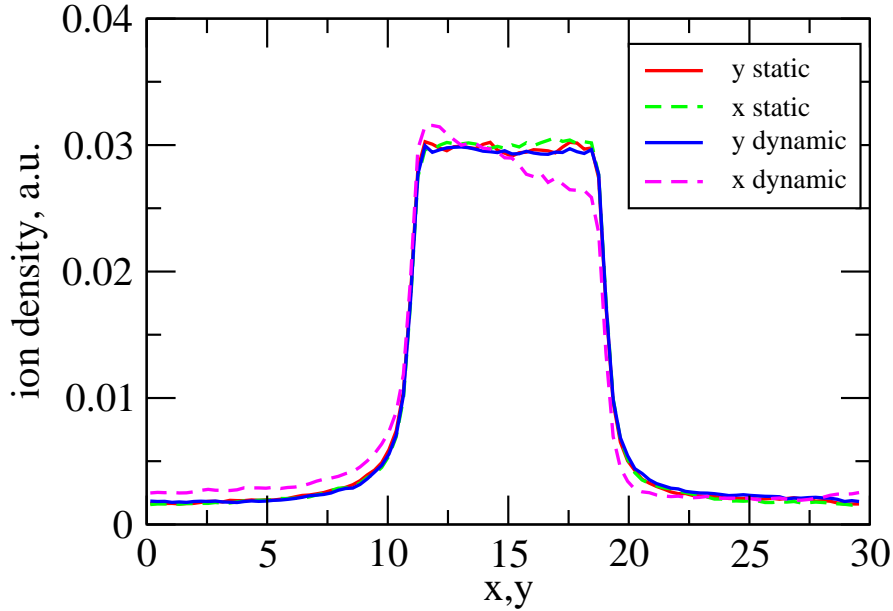}
\end{center}
\caption{Ion number density profile for a colloid 
  of charge -120 (i) in static conditions (solid lines) and (ii)
  in an external electric field of absolute value $E=0.2$
  so that the particle moves in positive $x$ direction (dashed lines). 
  The density profiles are taken in $x$ direction (along the
  field, $y$ and $z$ dependence integrated out) and
  in $y$ direction (perpendicular to the field, $x$ and $z$ dependence
  integrated out). The peak on the left hand side indicates 
  ion accumulation behind the particle.}
\label{fig:profile}
\end{figure}

\begin{figure}
\vskip 0.1in
\begin{center}
\includegraphics[clip, width=0.4\textwidth]{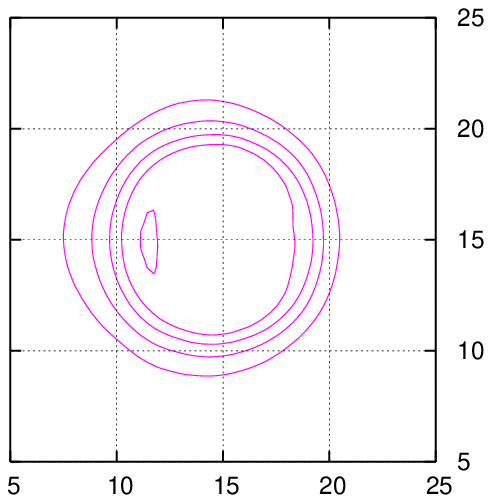}
\includegraphics[clip, width=0.4\textwidth]{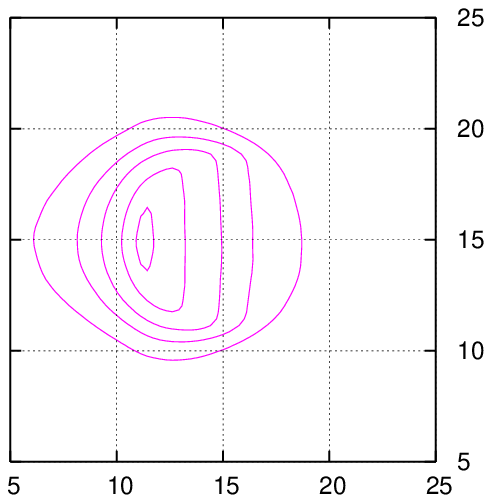}
\end{center}
\caption{Contour plots of the ion density distribution for a colloid 
  of charge -60 in an external electric field $E=0.2$ (left) and
  $E=0.6$ (right). The particle moves in the positive $x$ direction
  and its center is located at [15,15], while $\sigma_{ci} = 4$. 
  For this plot, we have only retained the $x$ and $y$ dependence
  of the density, while the $z$ dependence has been integrated out.
  The isolines are stretched to the left hand side,
  which indicates ion accumulation behind the particle.}
\label{fig:cont}
\end{figure}

In a stationary state, the motion of the ions in the double layer is
controlled by a balance of the external homogeneous force field, the
attraction to the colloid surface, the ion-ion interaction, and the
friction force. Thus, the stationary state double layer around the
colloid is in general asymmetric. The density profile along the field
(Fig. \ref{fig:profile}, dashed curves) does not coincide any longer
with that taken across the field direction (solid curves). One can
notice an accumulation of ions behind the particle (the particle is
moving towards larger $x$) and a rarefied region just in front of the
particle. Note however that the relative difference is not large,
which means that the ionic cloud is not strongly polarized.
Contour plots of the ionic distribution for a colloid with $Z=-60$,
$E=0.2$ and $E=0.6$ are shown in Fig. \ref{fig:cont}. The asymmetry of
the ionic cloud is not very large for these fields, which allows us to
regard the ionic cloud as spherical, for example in determining the effective
charges and potentials in a simple way.

\begin{figure}
\vskip 0.1in
\begin{center}
\includegraphics[clip, width=0.75\textwidth]{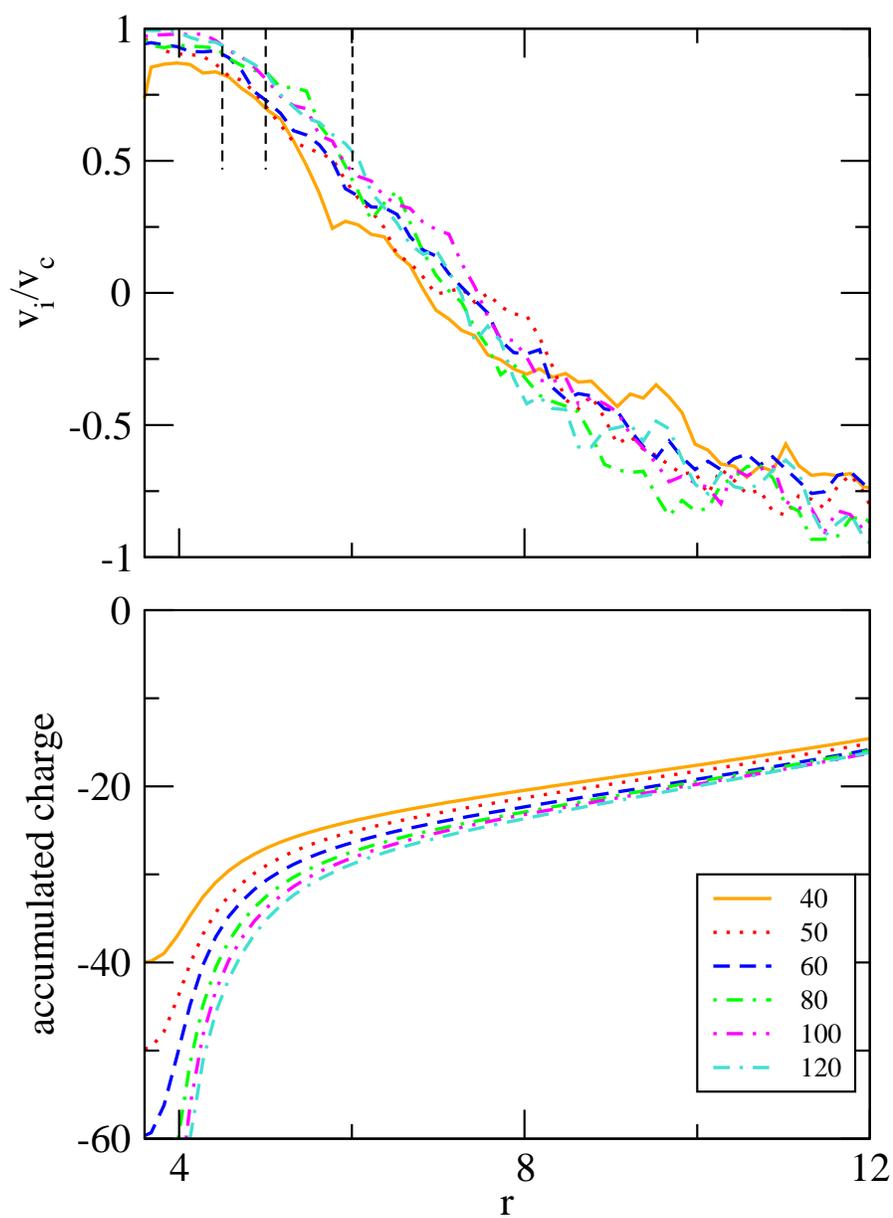}
\end{center}
\caption{Top: Radial distribution of the ionic drift velocity along the
  external field direction for colloids of different charge in an
  external electric field $E=0.2$. The thin dashed lines indicate the three 
  supposed positions of the slip surface. Bottom: Integrated charge curves
  for the same samples. The curves in both plots are marked by the
  bare colloid charge.}
\label{fig:zeta2}
\end{figure}

In order to define a dynamic effective charge of the colloid, which is
related to the zeta potential, we look at the radial distribution of
the ionic velocity in the double layer, taking the component in the
direction of the external field. As the instant velocities are
governed by thermal fluctuations and thus do not carry a sufficient
amount of the interesting information, we calculated the average
velocity over 10 LJ time units. This procedure averages out the
stochastic components of the velocity so that the directional drift
dominates the motion. The corresponding average curves are shown in
Fig. \ref{fig:zeta2}. The velocity is shown relative to the mean
velocity of the colloid. We see that the ions in the nearest
surroundings of the particle surface move along with the colloid. The
relative velocity reaches unity at about $r=4$, which agrees with the
peak of the ionic radial distribution. From there, the correlation
decreases monotonously and turns negative at about $r=8$. We then see
a region of anti-correlation spanning until $r=15$, the half-box
distance. In this region, the ionic drift velocity is anti-parallel to
the colloidal one. Since there is no well-defined plateau near the
particle, it is not possible to define a slip surface (where the ion
motion stops to be fully correlated with that of the central particle)
unambiguously, and thus we have taken several reasonable values for
the corresponding radius ($r = 4.5$, $r = 5.0$, $r = 6.0$). This
defines, on the one hand, the effective hydrodynamic radius of the
overall object, and on the other hand the effective charge (i.~e. the
charge within the slip surface).  Furthermore, the zeta potential
would be the electrostatic potential at this surface.

\begin{figure}
\vskip 0.1in
\begin{center}
\includegraphics[clip, width=0.75\textwidth]{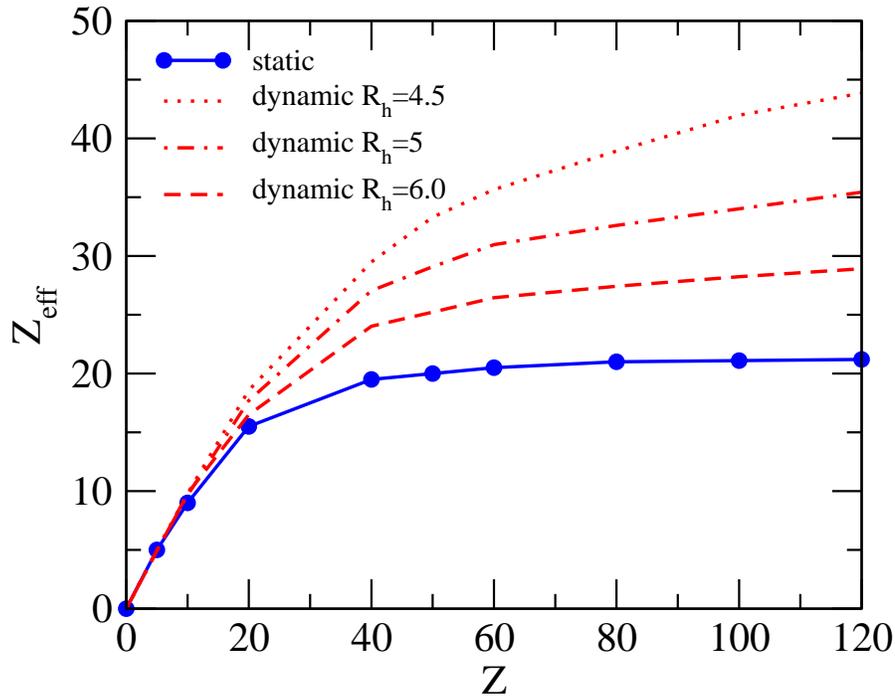}
\end{center}
\caption{ Static and dynamic effective charge of a colloidal particle as a 
  function of its structural charge. The static charge is calculated using
  a static simulation and the inflection point criterion. The dynamic charge
  is found using integration of the stationary charge distribution within the
  assumed shear surface at $E=0.2$. The corresponding colloid volume
  fraction is about 1\%.}
\label{fig:qsqd}
\end{figure}

We now turn to the relation between the static and dynamic effective
charges of a colloidal particle. This question reflects one of the
most common problems for technological applications of colloidal
dispersions: Predicting the stability of a suspension based on
electrophoretic mobility data or vice versa. Extensive studies have
been dedicated to the interpretation of electrokinetic data in order
to use them for particle characterization \cite{Dukhin,Hunter}. The
discrepancy between the two types of effective charges has a quite
obvious origin: The static and dynamic experimental setups probe
different properties of the ionic double layer. While static
properties like structure factor, elastic moduli, phase diagrams are
affected by the long-range decay of the potential, the particle
mobility depends primarily on the resistance of the double layer to
the viscous drag force. In both cases, it is not the particle surface
charge (or bare charge) that enters the definitions of the relevant
electrostatic and electrokinetic potentials but some effective value,
which is usually lower than the bare charge. Normally, it is difficult
to access both the static and the dynamic properties of the same
system and such a work requires a special effort. Recently, a
comparison between the static and dynamic effective charges calculated
from crystal shear moduli and suspension conductivity, respectively,
was done \cite{Palberg}. The dynamic effective charge was found to be
40\% larger than the static one for several sets of silica and latex  
particles in law-salt conditions. We expect that our system without
added salt shows similar features as the hydrodynamic coupling of the
ions to the particle is a very short-range effect.

The colloid-ion rdf's were integrated at vanishing external field to
obtain accumulated charge curves and to determine the static effective
charge by means of the inflection point criterion \cite{Belloni,Holm}.
Since it is known that most properties (in particular, the structure
factor) are not particularly sensitive to the precise definition of
the static charge (several different criteria turn out to be roughly
equivalent) \cite{Belloni}, we have just taken this for
convenience. For charges smaller that 20, no inflection point could be
detected and we took the effective charge at the inflection point
location of the more strongly charged particles. In the dynamic case,
the external field $E=0.2$ was applied. The radial charge
distributions were integrated in the stationary state, however not up
to the inflection point but rather up to the supposed slip surface, taking 
the three different values $r = 4.5$, $5.0$, and $6.0$ (see discussion
above, and Fig. \ref{fig:zeta2})

The resulting charge curves are shown in Fig. \ref{fig:qsqd}. Both the
static and dynamic effective charge grow linearly at $Z \rightarrow 0$
and more slowly at $Z>40$. The static charge curve saturates at a
value of $Z_{eff}^s \approx 22 $, while the dynamic charges continue
to increase. The charge calculated with the largest slip surface
radius $r=6.0$ reaches a magnitude of about 29, which is 30\% higher
than the static charge value. The two others, $r=4.5$ and $r=5.0$,
stop at 40.5 and 35.5, respectively. The observed difference between
the static and dynamic charges is by no means surprising as the
reasonable radii of the slip surface are all smaller than the
inflection point position, $r \approx 8$, and hence include less
counterions. Let us recall that the correlation between the colloid
and ion velocity in the dynamic measurement vanishes at this distance
$r = 8$ (see Fig. \ref{fig:zeta2}). In our case of no salt the
relation between the static and dynamic charges is indeed similar to
what has been reported for latex and silica spheres \cite{Palberg},
although we measured the static and dynamic charges in a different
way. We should note that this ratio is obviously not universal and
should hold only for electrokinetically similar systems. We expect it
to be much closer to unity in strongly screened systems with high salt
content as the double layer thickness would be smaller and thus closer
to the slip surface.

\begin{figure}
\vskip 0.1in
\begin{center}
\includegraphics[clip, width=0.75\textwidth]{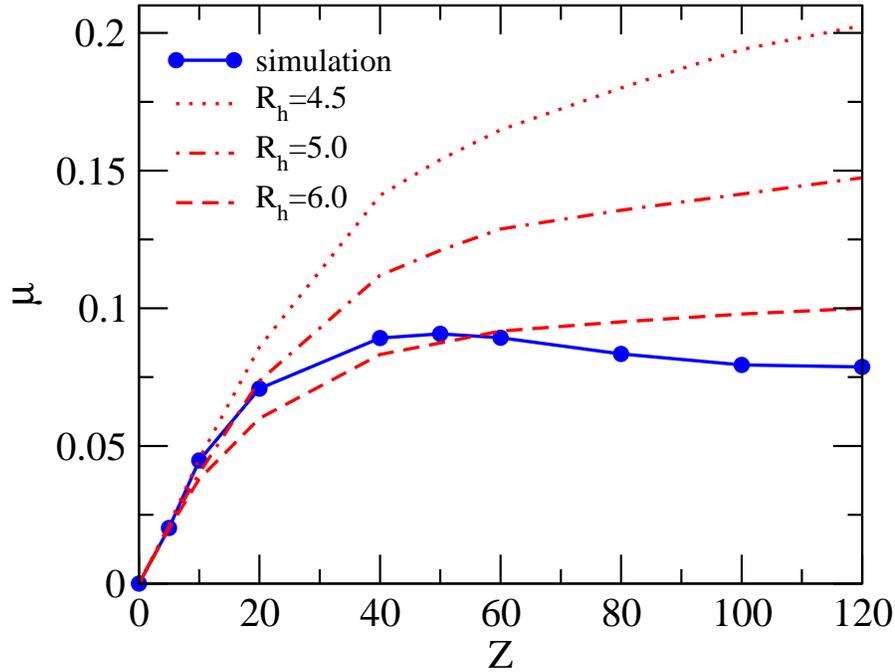}
\end{center}
\caption{Electrophoretic mobility of a particle of charge $Z$ and radius
  $R=3$ at external field $E=0.2$ in a simulation box of length 30 as  
  obtained directly from simulation (solid curve) or calculated from  
  Eq. \ref{eq:mu} with the indicated slip surface position and $f(\kappa
  R)=1$: $\mu=Z^d_{eff}e/(6 \pi \eta R_h)$. }
\label{fig:eph_q}
\end{figure}

We also looked at the particle electrophoretic mobility as a function
of particle bare charge. The considered charges range from 0 to
120. The electric field was set to a compromise value of 0.2 to make
the particle drift pronounced but to avoid strong asymmetry of the
double layer (see Fig. \ref{fig:cont}). The result is displayed in
Fig. \ref{fig:eph_q}. At low charges $Z \leq 10$, the mobility
increases with the charge almost linearly. This behavior can be
readily explained by a plain increase of the net force acting on the
particle. At very low charge, the interaction between the colloid and
the ions does not exceed $k_BT$, which makes the ion distribution only
slightly correlated with the colloid. At higher charges however the
increase of the electrophoretic mobility is slowing down, marking the
onset of ion condensation on the colloid surface. At $Z \approx 50$,
the mobility shows a maximum. Finally, at $Z>60$, we see a slight
decrease in the mobility upon further charge increase. This behavior
can be explained by the combined effect of colloid charge saturation
and increase of the friction due to the more and more packed double
layer. A similar behavior was reported already back in seventies for
charged emulsion droplets \cite{Dukhin}. A calculation using
Eq. \ref{eq:mu} with the preselected fixed slip surface positions and
$f(\kappa R)=1$ fails to reproduce this feature. The calculated
mobilities shown in Fig. \ref{fig:eph_q} (dashed or dotted curves)
grow monotonically.  One can see however that the curve with the
smaller $R_h=4.5$ describes better the initial part of the measured
mobility curve while the final part is described better by
$R_h=6$. Obviously, a consistent description requires that the
position of the slip surface (the hydrodynamic radius) shifts to
larger distances with increasing bare particle charge.

\begin{figure}
\vskip 0.1in
\begin{center}
\includegraphics[clip,width=0.75\textwidth]{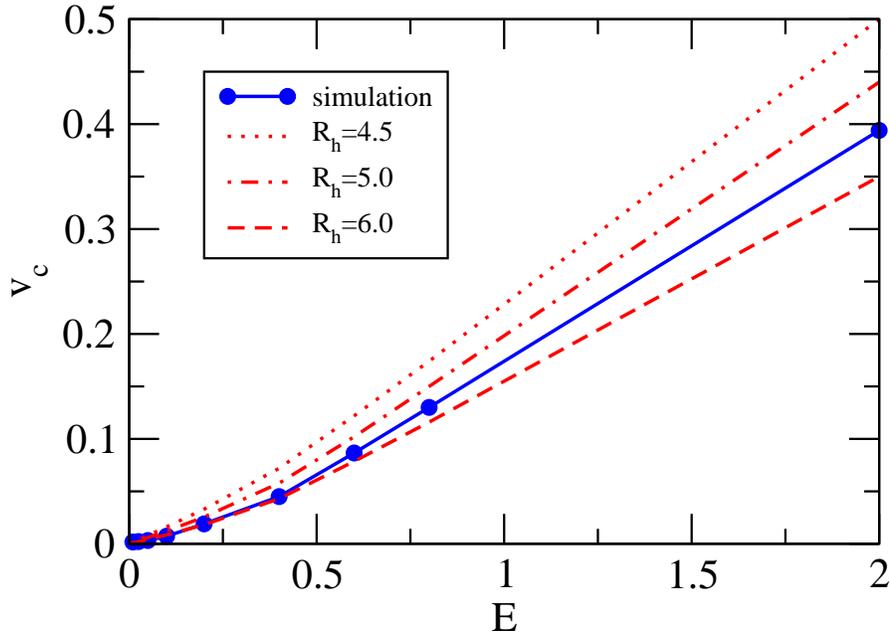}
\end{center}
\caption{Stationary drift velocity of a colloidal particle 
  with charge -60 in an external electric field $E$ as obtained from
  simulation or calculated using Eq. \ref{eq:1}-\ref{eq:mu} with the 
  indicated slip surface position and $f(\kappa R)=1$: $v=
  Z^d_{eff}e E/(6 \pi \eta R_h)$ }
\label{fig:eph_e}
\end{figure}

We finally looked at the colloid velocity as a function of the
external field (Fig. \ref{fig:eph_e}). In this way one can estimate
the limits of validity of linear response theory. In a weak field up
to $E \approx 0.5$, the velocity grows linearly (solid line) as
expected for a system with a constant electrophoretic mobility. In a
stronger field however a faster growth is seen, which means that the
mobility increases. We again supposed the three different slip surface
positions to calculate the velocity. A calculation with the shortest
distance $r=4.5$ predicts a too high velocity for all field
values. The curves corresponding to $r=5$ and $r=6$ bracket the
velocity values obtained by direct measurement in the simulation. The
best prediction can hence be made with $r \approx 5.5$. The most
important observations in this test are as follows: We found that 
for large field values $E>0.5$ no pronounced correlation of
the ion velocity with the colloid velocity was observed. The ratio
$v_i/v_c$ did not exceed 0.4 for $E=2$. Thus, the slip surface could
not be defined at all. This fact is supposedly related to the
distortion of the ionic cloud, which was illustrated in
Fig. \ref{fig:cont}. One should remember that the slip surface has to be 
defined by the ionic distributions in the limit $E \to 0$. Nevertheless, 
a reasonable estimate of the drift velocity can be obtained using the 
slip surface position found in the linear regime at small but finite $E$.

\section{Summary}
\label{sec:summary}

We applied a hybrid MD/LB simulation method for studying colloidal
electrophoresis. A combination of the primitive electrolyte model with
accurate treatment of the hydrodynamics allows us to access
experimentally measurable quantities like the electrophoretic mobility
and the underlying double layer structure at the same time. Our
simulation model shows the non-trivial coupling between charge
distribution and hydrodynamic flow, and the corresponding non-linear
effects. By analyzing the stationary ionic velocity distribution
around the colloid, we attempted to find the position of the slip
surface and the corresponding dynamic effective charge. Obviously, the
data for our model do not permit a unique definition; nevertheless,
from a combined analysis of the distribution, and the particle
mobility in terms of the H\"uckel theory, we could find reasonable
values. The dynamic charge appears to be somewhat larger than the
static effective charge estimated from the inflection point criterion,
which agrees with experimental findings for charged silica and latex
colloidal spheres. Our statistical definition of the slip surface
combined with the H\"uckel theory of electrophoresis gives a
reasonable description of the particle mobility in weak electric
fields in a salt-free suspension. A more detailed comparison with 
existing theories as well as with multi-colloid simulations will 
(hopefully) provide a deeper understanding of these issues in the future.

\ack We thank J\"urgen Horbach, Olga Vinogradova, and Kurt Kremer for
stimulating discussions. This work was funded by the SFB TR 6 of the
Deutsche Forschungsgemeinschaft.

\end{document}